\documentclass[preprint,amsmath,showpacs]{revtex4}
\usepackage{graphicx}
\usepackage{dcolumn}
\usepackage{bm}
\usepackage{epstopdf}

\def\bee{\begin{eqnarray}}
\def\eee{\end{eqnarray}}

\begin{document}
\draft
\title{Determination of the Neutrino Flavor Ratio at the Astrophysical Source  }
\author{Kwang-Chang Lai, Guey-Lin Lin and T. C. Liu}\affiliation{Institute of Physics, National
Chiao-Tung University, Hsinchu 300, Taiwan}\affiliation{Leung Center
for Cosmology and Particle Astrophysics, National Taiwan University,
Taipei 106, Taiwan.}

\date{\today}

\begin{abstract}
We discuss the reconstruction of neutrino flavor ratios at
astrophysical sources through the future neutrino-telescope
measurements. Taking the ranges of neutrino mixing parameters
$\theta_{ij}$ as those given by the current global fit, we
demonstrate by a statistical method that the accuracies in the
measurements of energy-independent ratios $R\equiv\phi
(\nu_{\mu})/\left(\phi (\nu_{e})+\phi (\nu_{\tau})\right)$ and
$S\equiv\phi (\nu_e)/\phi (\nu_{\tau})$ among integrated neutrino
flux should both be better than $10\%$ in order to distinguish
between the pion source and the muon-damped source at the $3\,
\sigma$ level. The $10\%$ accuracy needed for measuring $R$ and $S$
requires an improved understanding on the background atmospheric
neutrino flux to a better than $10\%$ level in the future. We
discuss the applicability of our analysis to practical situations
that the diffuse astrophysical neutrino flux arises from different
types of sources and each point source has a neutrino flavor ratio
varying with energies. We also discuss the effect of leptonic CP
phase on the flavor-ratio reconstruction.
\end{abstract}

\maketitle

\section{Introduction}
The operation of IceCube detector \cite{Berghaus:2008bk} and the
R\&D effort of KM3Net \cite{km3net} are important progresses toward
a km$^3$-sized detection capability in the neutrino astronomy
\cite{Berezinsky:2009zw}. Furthermore the radio and air-shower
detectors, such as ANITA \cite{Gorham:2008yk} and Pierre Auger
detector \cite{Collaboration:2009uy} respectively, are also taking
the data. These detectors are sensitive to neutrinos with energies
higher than those probed by IceCube and KM3Net. Finally, the radio
extension of IceCube detector, the IceRay \cite{Allison:2009rz}, is
also under consideration. It is expected to detect a score of
cosmogenic neutrinos \cite{cosmo_nu} per year. Motivated by the
development of neutrino telescopes, numerous  efforts were devoted
to studying neutrino mixing parameters with astrophysical neutrinos
as the beam source
\cite{Beacom:2003zg,Pakvasa:2004hu,Costantini:2004ap,
Bhattacharjee:2005nh,
Serpico:2005sz,Serpico:2005bs,Xing:2006uk,Winter:2006ce,Xing:2006xd,
Majumdar:2006px,Rodejohann:2006qq,Meloni:2006gv,Blum:2007ie,Hwang:2007na,
Pakvasa:2007dc,Choubey:2008di,Maltoni:2008jr}. Due to the large
neutrino propagation distance, the neutrino oscillation
probabilities only depend on the mixing angles $\theta_{ij}$ and the
CP phase $\delta$ \cite{Learned:1994wg,AB}, which make the
astrophysical beam source favorable for extracting the above
parameters, provided there are sufficient number of events.

Most of the astrophysical neutrinos are believed to be produced by
the decay of charged pion through the following chain: $\pi^+\to
\mu^+ +\nu_{\mu}\to e^+ +\nu_{\mu}+\nu_e+\bar{\nu}_{\mu}$ or
$\pi^-\to \mu^- +\bar{\nu}_{\mu}\to e^-
+\bar{\nu}_{\mu}+\bar{\nu}_e+\nu_{\mu}$. This leads to the neutrino
flux ratio $ \phi_0(\nu_e): \phi_0(\nu_{\mu}):
\phi_0(\nu_{\tau})=1:2:0$ at the astrophysical source where
$\phi_0(\nu_{\alpha})$ is the sum of $\nu_{\alpha}$ and
$\bar{\nu}_{\alpha}$ flux. Such a flux ratio results from an
implicit assumption that the muon decays into neutrinos before it
loses a significant fraction of its energy. However, in some source
the muon quickly loses its energy by interacting with strong
magnetic fields or with matter
\cite{Rachen:1998fd,Kashti:2005qa,Kachelriess:2007tr}. Such a muon
eventually decays into neutrinos with energies much lower than that
of $\nu_{\mu} (\bar{\nu}_{\mu})$ from $\pi^+(\pi^-)$ decays.
Consequently this type of source has a neutrino flavor ratio  $
\phi_0(\nu_e): \phi_0(\nu_{\mu}): \phi_0(\nu_{\tau})=0:1:0$, which
is referred to as the muon-damped source. The third type of source
emits neutrons resulting from the photo-disassociation of nuclei. As
neutrons propagate to the Earth, $\bar{\nu}_e$ are produced from
neutron $\beta$ decays \cite{Anchordoqui:2003vc}, leading to a
neutrino flavor ratio $\phi_0(\nu_e): \phi_0(\nu_{\mu}):
\phi_0(\nu_{\tau})=1:0:0$. Finally, neutrinos might be produced deep
inside optically thick sources so that the flavor ratio at the
source surface is significantly different from the flavor ratio at
the production point due to the oscillations \cite{Mena:2006eq}.
Hence, unlike the previous three cases, the $\nu_{\tau}$ fraction
can be significant at the surface of such sources. In the class of
sources studied by Mena {\it et al.} \cite{Mena:2006eq}, which are
referred to as the astrophysical hidden sources, the neutrino flux
ratio for $E_{\nu}> 10^4$ GeV approaches to $1/3:a:b$ with both $a$
and $b$ oscillating with the neutrino energy under the constraint
$a+b=2/3$.

As mentioned before, almost all previous studies treat astrophysical
neutrinos as the beam source for extracting neutrino mixing
parameters \cite{note}. To have a better determination of certain
neutrino mixing parameter, for instance the atmospheric mixing angle
$\theta_{23}$ or the CP phase $\delta$, a combined analysis on the
terrestrially measured flavor ratios of astrophysical neutrinos
coming from different sources, such as the pion source and the
muon-damped source, has been considered
\cite{Blum:2007ie,Choubey:2008di}. A natural question to ask is then
how well one can distinguish these neutrino sources. The answer to
this question depends on our knowledge of neutrino mixing parameters
and the achievable accuracies in measuring the neutrino flavor
ratios on the Earth such as $R\equiv\phi (\nu_{\mu})/\left(\phi
(\nu_{e})+\phi (\nu_{\tau})\right)$ and $S\equiv \phi (\nu_e)/\phi
(\nu_{\tau})$. In this article, we shall provide an answer to this
question with a statistical analysis.

The possibility of measuring neutrino flavor fraction by IceCube has
been discussed in Ref.~\cite{Beacom:2003nh}. It is through the
measurement of muon track to shower ratio. It was demonstrated that
the $\nu_e$ fraction can be extracted from the above ratio by
assuming flavor independence of the neutrino spectrum and the
$\nu_{\mu}-\nu_{\tau}$ symmetry, i.e., $\phi (\nu_{\mu})=\phi
(\nu_{\tau})$. Taking a pion source with $E^2\phi (\nu_{\mu})=
10^{-7}$ GeV cm$^{-2}$s$^{-1}$ \cite{Waxman:1998yy} and thresholds
for muon and shower energies taken to be $100$ GeV and $1$ TeV
respectively, the $\nu_e$ fraction can be determined to an accuracy
of $25\%$ at IceCube for one year of data taking, or equivalently to
an accuracy of $8\%$ for a decade of data taking. However, the tau
neutrino events are too rare to provide additional information on
the neutrino flavor composition. The analysis in
Ref.~\cite{Beacom:2003nh} as summarized above provides a feasibility
of measuring $R$ in a good precision at IceCube and detectors with
comparable capacities. In fact one may repeat the analysis in
\cite{Beacom:2003nh} and extract $R$ and its associated uncertainty
directly. The uncertainty in $R$ is expected to be comparable to
that of $\nu_e$ fraction. We note that the precisions on measuring
$R$ and $S$ should depend on neutrino energies. However, for
simplicity in discussions, we shall take $R$ and $S$ as ratios of
integrated neutrino flux with appropriate energy thresholds for
suppressing the atmospheric neutrino background. These ratios and
the corresponding precisions, $\Delta R/R$ and $\Delta S/S$, are
therefore energy independent. In our analysis, we do not assume
$\nu_{\mu}-\nu_{\tau}$ symmetry for the neutrino flux measured on
the Earth. We shall argue that, besides measuring $R$, it is
essential to measure $S$ in order to reconstruct the neutrino flavor
ratio at the astrophysical source. This implies that a neutrino
telescope beyond the capability of IceCube is needed to study the
neutrino flavor astronomy.

It is important to understand the atmospheric neutrino background
which affects the precisions on measuring $R$ and $S$. The flux
spectrum of conventional atmospheric neutrinos which arise from pion
and kaon decays is well understood
\cite{Gaisser:2002jj,Barr:2006it}. The measurement on such a
spectrum \cite{Collaboration:2009nf} has reached to the energy of
$10^5$ GeV. The prompt atmospheric neutrino flux arising from charm
decays still contains large uncertainties
\cite{Costa:2001fb,Bugaev:1998bi,Martin:2003us} and it has not yet
been measured experimentally. The prompt atmospheric neutrino flux
takes over the conventional one around $10^5$ GeV for $\nu_e$ and
$10^6$ GeV for $\nu_{\mu}$ \cite{BA}. The flavor ratio of
conventional atmospheric neutrino beyond TeV energies is
approximately $ \phi_c^{\rm atm}(\nu_e): \phi_c^{\rm
atm}(\nu_{\mu}): \phi_c^{\rm atm}(\nu_{\tau})=1:20:0$; while the
flavor ratio of prompt atmospheric neutrino flux is approximately
$\phi_p^{\rm atm}(\nu_e): \phi_p^{\rm atm}(\nu_{\mu}): \phi_p^{\rm
atm}(\nu_{\tau})=1:1:0.1$. Such flavor ratios differ significantly
from those of astrophysical neutrinos which arrive on Earth with
$\phi(\nu_{\mu})\approx \phi(\nu_{\tau})$. To suppress atmospheric
neutrino background in the search of astrophysical neutrinos, energy
distributions of astrophysical neutrino events and cuts on PMT hits
are imposed \cite{AMANDAII}.

The paper is organized as follows. In Sec. II, we discuss properties
of the probability matrix that links the initial neutrino flavor
ratio to the ratio measured on the Earth. In Sec. III, we begin with
a brief review on the current understanding of neutrino mixing
angles. We then present the reconstructed neutrino flavor ratio at
the source from the simulated data, which is generated by the chosen
true values of the neutrino flavor ratio at the source and best-fit
values of neutrino mixing parameters. The statistical analysis is
performed with different measurement accuracies in $R$ and $S$, as
well as different ranges of neutrino mixing parameters. The
implications of our results are discussed in Sec. IV.

\section{Neutrino mixing parameters and oscillations of astrophysical neutrinos}\label{matrix}
The neutrino flux at the astrophysical source and that detected on the
Earth are related by
\begin{eqnarray}
\left(
  \begin{array}{c}
     \phi(\nu_e) \\
    \phi (\nu_{\mu}) \\
    \phi (\nu_{\tau})\\
  \end{array}
\right)
 =
\left(
   \begin{array}{ccc}
     P_{ee} & P_{e\mu} & P_{e\tau} \\
     P_{\mu e} & P_{\mu\mu} & P_{\mu\tau} \\
     P_{\tau e} & P_{\tau\mu} & P_{\tau\tau} \\
   \end{array}
 \right)
 \left(
  \begin{array}{c}
     \phi_0(\nu_e) \\
    \phi_0(\nu_{\mu}) \\
    \phi_0(\nu_{\tau})\\
  \end{array}
\right)\equiv P\left(
  \begin{array}{c}
     \phi_0(\nu_e) \\
    \phi_0(\nu_{\mu}) \\
    \phi_0(\nu_{\tau})\\
  \end{array}
\right)
, \label{source_earth}
\end{eqnarray}
where $\phi(\nu_{\alpha})$ is the neutrino flux measured on the
Earth while $\phi_0(\nu_{\alpha})$ is the neutrino flux at the
astrophysical source, and the matrix element $P_{\alpha\beta}$ is the probability of the oscillation
$\nu_{\beta}\to \nu_{\alpha}$. The exact analytic expressions for $P_{\alpha\beta}$ are given in Eq.~(\ref{exact_prob}). It is seen that $P_{e\mu}=P_{e\tau}$ and $P_{\mu\mu}=P_{\mu\tau}=P_{\tau\tau}$ in the limit $\Delta=0=D$, i.e., $\theta_{23}=\pi/4$ and $\theta_{13}=0$. In this case, the probability matrix $P$ is singular with a vanishing determinant. In general, the determinant of this matrix remains suppressed since both $\Delta$ and $D$ are expected to be small. For $\Delta=0=D$, the eigenvectors of $P$ are given by
\begin{eqnarray}
V^a=\frac{1}{\sqrt{3}}
      \left(
        \begin{array}{c}
          1 \\
          1 \\
          1 \\
        \end{array}
      \right)
,\, V^b=\frac{1}{\sqrt{2}}
\left(
  \begin{array}{c}
    0 \\
    -1 \\
    1 \\
  \end{array}
\right)
,\, V^c=\frac{1}{\sqrt{6}}
\left(
  \begin{array}{c}
    -2 \\
    1 \\
    1 \\
  \end{array}
\right)
,\label{vabc}
\end{eqnarray}
with the corresponding eigenvalues
\begin{equation}
\lambda_a=1,\, \lambda_b=0,\, \lambda_c=\frac{1}{4}(4-3\omega),
\end{equation}
where $\omega=\sin^2 2\theta_{12}$. Therefore, those initial flavor ratios that differ from one another by a multiple of $V^b$
shall oscillate into the same flavor ratio on the Earth. To illustrate this explicitly, we write the initial flux $\Phi_0$ at the astrophysical source as
\begin{equation}
\Phi_0=\left(
         \begin{array}{c}
           1 \\
           0 \\
           0 \\
         \end{array}
       \right)
       -\frac{\sqrt{2}}{2}\left(\phi_0 (\nu_{\mu})-\phi_0 (\nu_{\tau})\right)V^b
       +\frac{\sqrt{6}}{2}\left(\phi_0 (\nu_{\mu})+\phi_0 (\nu_{\tau})\right)V^c,\label{initial}
\end{equation}
where we have imposed the normalization condition $\phi_0 (\nu_e)+\phi_0
(\nu_{\mu})+\phi_0 (\nu_{\tau})=1$. This normalization convention will be adopted throughout this paper.
The first term on the right-hand side (RHS) of Eq.~(\ref{initial}) can be expressed as $(\sqrt{3}V^a-\sqrt{6}V^c)/3$. Hence the neutrino flux measured by the terrestrial neutrino telescope is
\begin{equation}
\Phi=P\phi_0=\frac{\sqrt{3}}{3}V^a-\frac{\sqrt{6}}{3}(1-\frac{3}{4}\omega)V^c +\frac{\sqrt{6}\lambda_c}{2}\left(\phi_0 (\nu_{\mu})+\phi_0 (\nu_{\tau})\right)V^c.
\end{equation}
It is seen that the vector $V^b$, with a coefficient proportional to $\phi_0 (\nu_{\mu})-\phi_0 (\nu_{\tau})$, does not appear in the terrestrially measured flux $\Phi$. Hence the terrestrial measurement can not constrain $\phi_0 (\nu_{\mu})-\phi_0 (\nu_{\tau})$ in this case.

The above degeneracy is lifted by either a non-vanishing
$\theta_{13}$ ($D\neq 0$) or a deviation of $\theta_{23}$ from $\pi/4$ ($\Delta\neq 0$). To simplify our discussions, let us take $D=0$ and $\Delta\neq 0$.
One can show that
the flux combination $\left(1+4\omega\Delta/(4-3\omega)\right)\phi_0 (\nu_{\mu})-\left(1-2\omega\Delta/(4-3\omega)\right)\phi_0 (\nu_{\tau})$ remains poorly constrained due to the suppression of $\det P$.
To demonstrate this, we observe that
\begin{eqnarray}
P=\frac{1}{8} \left(
  \begin{array}{ccc}
    8-4\omega & 2(1+\Delta)\omega & 2(1-\Delta)\omega \\
    2(1+\Delta)\omega & (4-\omega)(1+\Delta^2)-2\Delta\omega & (4-\omega)(1-\Delta^2) \\
     2(1-\Delta)\omega &  (4-\omega)(1-\Delta^2) &  (4-\omega)(1+\Delta^2)+2\Delta\omega \\
  \end{array}
\right)
\end{eqnarray}
for $D=0$ and $\Delta\neq 0$.
The eigenvalues of $P$ expanded
to the second order in $\Delta$ are given by
\begin{equation}
\lambda^{\prime}_a=1,\,
\lambda^{\prime}_b=\left(\frac{4-4\omega}{4-3\omega}\right)\Delta^2,\,
\lambda^{\prime}_c=\frac{1}{4}(4-3\omega)+\frac{3\omega^2\Delta^2}{4(4-3\omega)},
\label{lapabc}
\end{equation}
and the corresponding eigenvectors to the same order in $\Delta$
are
\begin{eqnarray}
V^{\prime a} &=&
N^a\left(
  \begin{array}{c}
    1 \\
    1 \\
    1 \\
  \end{array}
\right)
,
\nonumber \\
V^{\prime b} &=&
N^b\left(
  \begin{array}{c}
    2r\Delta\left(1+r\Delta\right) \\
     -1-2r\Delta\left(1+r\Delta\right)\\
    1 \\
  \end{array}
\right)
, \nonumber \\
V^{\prime c}&=&
N^c\left(
  \begin{array}{c}
     -2+6r\Delta \\
    1-6r\Delta\left(1-3r\Delta\right) \\
     1\\
  \end{array}
\right),\label{vpabc}
\end{eqnarray}
with $r=\omega/(4-3\omega)$ and $N^{a,b,c}$ the appropriate
normalization factors. It is interesting to note that the
corrections to the eigenvectors of $P$ begin at $\mathcal{O}(\Delta)$
while the corrections to the corresponding eigenvalues begin at
$\mathcal{O}(\Delta^2)$. With the above eigenvectors, we write the source
neutrino flux as
 \begin{eqnarray}
\Phi_0 &=&N^aV^{\prime a}
       -\left[\left(1+4r\Delta\right)\phi_0 (\nu_{\mu})-
       \left(1-2r\Delta\right)\phi_0 (\nu_{\tau})-2r\Delta\right]N^bV^{\prime b}\nonumber \\
       &+&3\left[\left(1-4r\Delta\right)\phi_0 (\nu_{\mu})+
       \left(1-2r\Delta\right)\phi_0 (\nu_{\tau})-
       \frac{2}{3}\left(1-3r\Delta\right)\right]N^cV^{\prime c}.\label{initial_2}
\end{eqnarray}
 It is easy to show that the measured flux $P\Phi_0$ depends on $V^{\prime b}$
 through the combination $-B\lambda^{\prime}_bN^bV^{\prime b}$ with
\begin{equation}
B=\left[\left(1+4r\Delta\right)\phi_0
(\nu_{\mu})-\left(1-2r\Delta\right)\phi_0
(\nu_{\tau})-2r\Delta\right].
\end{equation}
 Clearly the flux combination
 $\left(1+4r\Delta\right)\phi_0 (\nu_{\mu})-
 \left(1-2r\Delta\right)\phi_0 (\nu_{\tau})$  is poorly constrained due to
 the smallness of $\lambda^{\prime}_b$, of the order $\Delta^2$.
\section{Statistical Analysis}

To reconstruct the neutrino flavor ratio at the source with a statistical analysis,
we employ the following best-fit values and $1\sigma$ ranges of neutrino mixing
angles \cite{GonzalezGarcia:2007ib}
\begin{eqnarray}
\sin^2\theta_{12}=0.32^{+0.02}_{-0.02}, \,
\sin^2\theta_{23}=0.45^{+0.09}_{-0.06}, \, \sin^2\theta_{13}< 0.019,
\label{bestfit}
\end{eqnarray}
for the major part of our analysis. In the above parameter set, the
best-fit value of $\theta_{23}$ is smaller than $\pi/4$. There exist
proposals to probe $\sin^2\theta_{23}$ by future atmospheric
neutrino experiments \cite{GonzalezGarcia:2004cu,Choubey:2005zy} and
long baseline neutrino experiments \cite{Huber:2004ug}. We therefore
include in our analysis the hypothetical scenario that
$(\sin^2\theta_{23})_{\rm best \, fit}=0.55$ with an error identical
to the one associated with $(\sin^2\theta_{23})_{\rm best \,
fit}=0.45$. Finally we also consider a $\theta_{13}$ range suggested
by Ref.~\cite{Fogli:2008jx} where
\begin{eqnarray}
\sin^2\theta_{13}=0.016\pm 0.010 (1\, \sigma)
\end{eqnarray}
by a global analysis.

In this work, we investigate uncertainties in the reconstruction of neutrino
flavor ratios at the source for the pion source and the muon-damped source.
Different choices of neutrino mixing parameters in our analysis are listed in Table \ref{scen}.
\begin{table}
  \centering
  \caption{Parameter sets chosen for our analysis}\label{scen}
  \begin{tabular}{ccccc}
\hline
       Parameter set &  $\sin^2\theta_{12}$  &  $\sin^2\theta_{23}$              &  $\sin^2\theta_{13}$  & $\delta$    \\
             1    & $0.32^{+0.02}_{-0.02}$     &   $0.45^{+0.09}_{-0.06}$      &  $<0.019$                        &     $0$   \\
             2    & $0.32^{+0.02}_{-0.02}$    &   $0.55^{+0.09}_{-0.06}$      &  $<0.019$                        &     $0$   \\
             3a    & $0.32^{+0.02}_{-0.02}$    &   $0.45^{+0.09}_{-0.06}$      &  $0.016\pm 0.010$               &     $0$   \\
              3b   & $0.32^{+0.02}_{-0.02}$     &   $0.45^{+0.09}_{-0.06}$      &  $0.016\pm 0.010$               &
              $\pi/2$   \\
              3c   & $0.32^{+0.02}_{-0.02}$     &   $0.45^{+0.09}_{-0.06}$      &  $0.016\pm 0.010$               &     $\pi$   \\
 \hline
\end{tabular}
\end{table}
Employing these mixing parameters, the true values of neutrino flavor ratios on the Earth and the corresponding values for $R$ and $S$ are presented in Table \ref{truevalue}. The true values of the neutrino flavor ratios on the Earth are denoted by $\Phi_{\pi}$ and $\Phi_{\mu}$
for the pion source and the muon-damped source respectively.
They are calculated with Eq.~(\ref{source_earth})
where $P$ is evaluated with neutrino mixing parameters at their best-fit values.
The flux ratios $R_{\pi}$ and $S_{\pi}$ are obtained from $\Phi_{\pi}$
while $R_{\mu}$ and $S_{\mu}$ are obtained from $\Phi_{\mu}$.
\begin{table}
  \centering
  \caption{True values of neutrino flavor ratios on the Earth}\label{truevalue}
  \begin{tabular}{ccccccc}
\hline
        Parameter set&   $\Phi_\mu=P\Phi_{0,\mu}$      &       $R_\mu$   &     $S_\mu$      &   $\Phi_\pi=P\Phi_{0,\pi}$     &     $R_\pi$     &     $S_\pi$ \\
             1         &   $(024,0.37,0.39)$                     &  $0.62$            &     $0.60$         &   $(0.35,0.33,0.32)$              &  $0.49$          &     $1.08$ \\
             2         &   $(0.19,0.42,0.39)$                    &  $0.71$            &     $0.51$         &   $(0.32,0.34,0.34)$               &  $0.52$          &     $0.94$\\
             3a     &   $(0.27,0.35,0.38)$                     &  $0.55$            &     $0.71$        &   $(0.36,0.33,0.31)$                &  $0.48$          &     $1.15$\\
             3b     &   $(0.25,0.37,0.38)$                     &  $0.59$            &     $0.64$        &   $(0.35,0.33,0.32)$                &  $0.49$          &     $1.07$\\
             3c     &   $(0.23,0.40,0.37)$                     &  $0.67$            &     $0.60$        &   $(0.33,0.34,0.33)$                &  $0.52$          &     $1.02$\\
 \hline
\end{tabular}
\end{table}

Given a precision on measuring $R$, $\Delta R_i/R_i$, we estimate
$\Delta S_i/S_i$ with two approaches. The first approach assumes
that both $\Delta R_i$ and $\Delta S_i$ are dominated by the
statistical errors. In this case, one has
\begin{eqnarray}
\left(\frac{\Delta
S_i}{S_i}\right)=\frac{1+S_i}{\sqrt{S_i}}\sqrt{\frac{R_i}{1+R_i}}\left(\frac{\Delta
R_i}{R_i}\right),
\end{eqnarray}
with $i=\pi,\, \mu$ \cite{Blum:2007ie}.  Using values of $R_i$ and
$S_i$ from Table \ref{truevalue}, we obtain  $\Delta
S_\pi/S_\pi=(1.1-1.2)(\Delta R_\pi/R_\pi)$ and $\Delta S_\mu/S_\mu=
(1.1-1.4)(\Delta R_\mu/R_\mu)$. The second approach takes into
account the specific complications for identifying tau neutrinos.
Since tau lepton decays before it loses a significant fraction of
its energy, tau neutrino is identified by the so-called double-bang
or lollipop events \cite{Learned:1994wg,Beacom:2003nh,Athar:2000rx}.
In IceCube or other detector with a comparable size, double-bang
events are observable only in a narrow energy range between $2$ PeV
and $20$ PeV \cite{Learned:1994wg,Athar:2000rx} while the
probability for observing a lollipop event, though increasing with
the neutrino energy, is still less than $10^{-3}$ for $E_{\nu}=1$
EeV \cite{Beacom:2003nh}. In view of these, we do not correlate
$\Delta S_i/S_i$ with $\Delta R_i/R_i$ in the second approach.
Rather we fix $\Delta S_i/S_i$ while vary $\Delta R_i/R_i$ for
achieving the goal of distinguishing astrophysical neutrino sources.
The results of both approaches will be presented. Before presenting
the details of our analysis, we point out that the decays $\tau \to
\nu_{\tau}\mu\bar{\nu}_{\mu}$ and $\tau \to
\nu_{\tau}\mu\bar{\nu}_{\mu}$, each with a $18\%$ branching ratio,
produce extra muon events or secondary $\nu_e$ and $\nu_{\mu}$
\cite{Halzen:1998be,Beacom:2001xn}. Cares are needed to separate
these events from those of primary $\nu_e$ and $\nu_{\mu}$ or muons
produced by the charged current interaction.

The fitting to the neutrino flavor ratios at the source is
facilitated through
\begin{equation}
\chi_i^2=\left(\frac{R_{i,\rm th}-R_{i,\rm exp}}{\sigma_{R_{i,\rm
exp}}}\right)^2+ \left(\frac{S_{i,\rm th}-S_{i,\rm
exp}}{\sigma_{S_{i,\rm exp}}}\right)^2 +\sum_{jk=12,23,13}
\left(\frac{s_{jk}^2-(s_{jk})^2_{\rm best \,
fit}}{\sigma_{s_{jk}^2}}\right)^2\label{chi}
\end{equation}
with $i=\pi,\, \mu$, $\sigma_{R_{i,\rm exp}}=(\Delta
R_i/R_i)R_{i,\rm exp}$, $\sigma_{S_{i,\rm exp}}=(\Delta
S_i/S_i)S_{i,\rm exp}$,  $s_{jk}^2\equiv \sin^2\theta_{jk}$ and
$\sigma_{s_{jk}^2}$ the $1\sigma$ range for $s_{jk}^2$. Here
$R_{i,\rm th}$ and $S_{i,\rm th}$ are theoretical predicted values
for $R_i$ and $S_i$ respectively while $R_{i,\rm exp}$ and $S_{i,\rm
exp}$ are experimentally measured values. The values for $R_{i,\rm
exp}$ and $S_{i,\rm exp}$ are listed in Table \ref{truevalue}, which
are generated from input true values of neutrino flavor ratios at
the source and input true values of neutrino mixing parameters. In
$R_{i,\rm th}$ and $S_{i,\rm th}$,  the variables $s_{jk}^2$ can
vary between $0$ and $1$ while $\cos \delta$ can vary between $-1$
and $1$. We note that similar $\chi^2$ functions have been used for
fitting the CP violation phase and the mixing angle $\theta_{23}$
respectively \cite{Blum:2007ie,Choubey:2008di}, assuming the source
flavor ratio is known. In our analysis, we scan all possible
neutrino flavor ratios at the source that give rise to a specific
$\chi_i^2$ value. Since we have taken $R_{i,\rm exp}$ and $S_{i,\rm
exp}$ as those generated by input true values of initial neutrino
flavor ratios and neutrino mixing parameters, we have
$(\chi_i^2)_{\rm min}$=0 occurring at these input true values of
parameters. Hence the boundaries for $1\sigma$ and $3\sigma$ ranges
of initial neutrino flavor ratios are given by $\Delta \chi_i^2=2.3$
and $\Delta \chi_i^2=11.8$ respectively where $\Delta \chi_i^2\equiv
\chi_i^2-(\chi_i^2)_{\rm min}=\chi_i^2$ in our analysis.

\subsection{The reconstruction of initial neutrino flavor ratio by measuring $R$ alone }
\begin{figure}[htbp]
\includegraphics[scale=0.35]{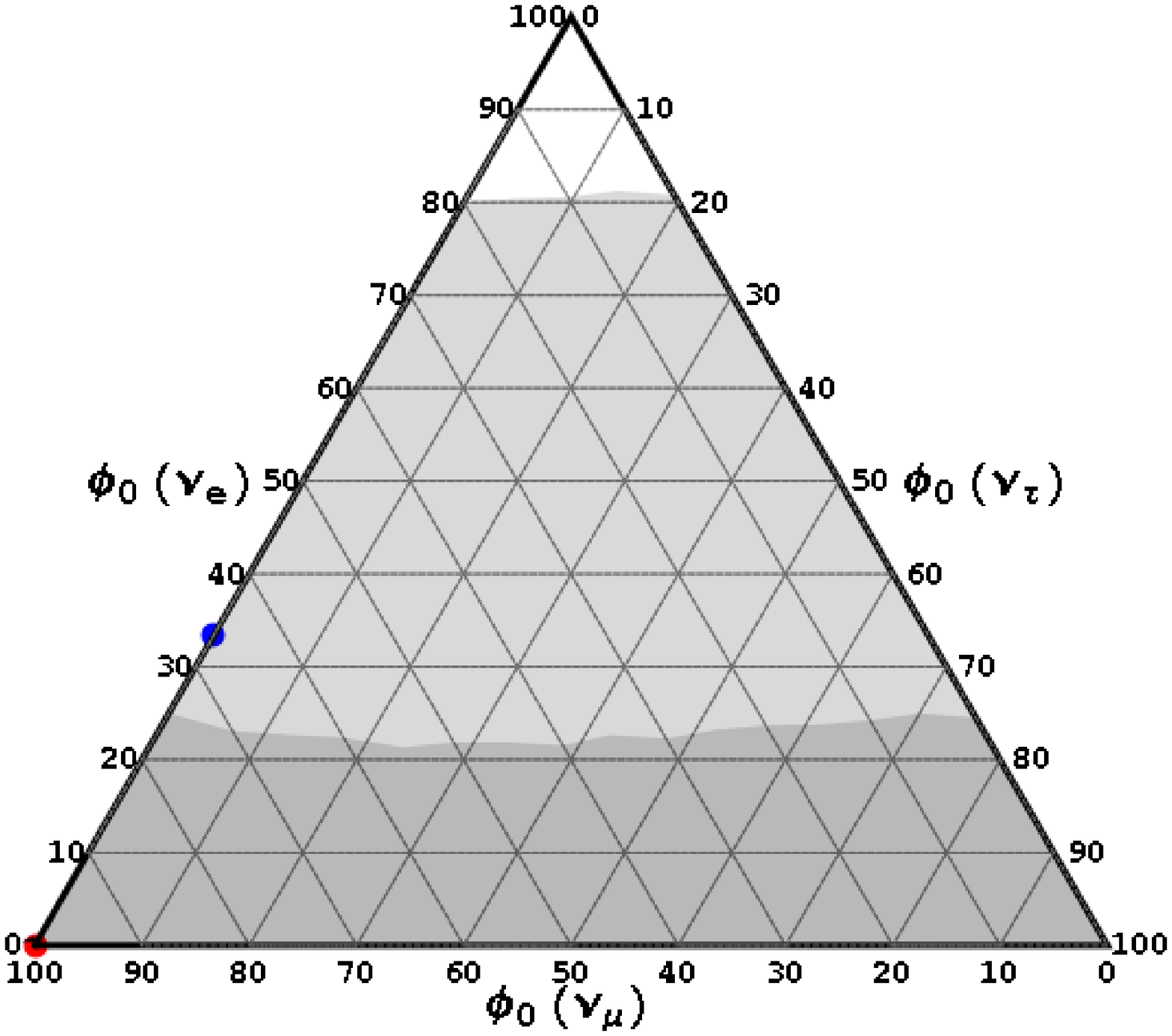}
\includegraphics[scale=0.35]{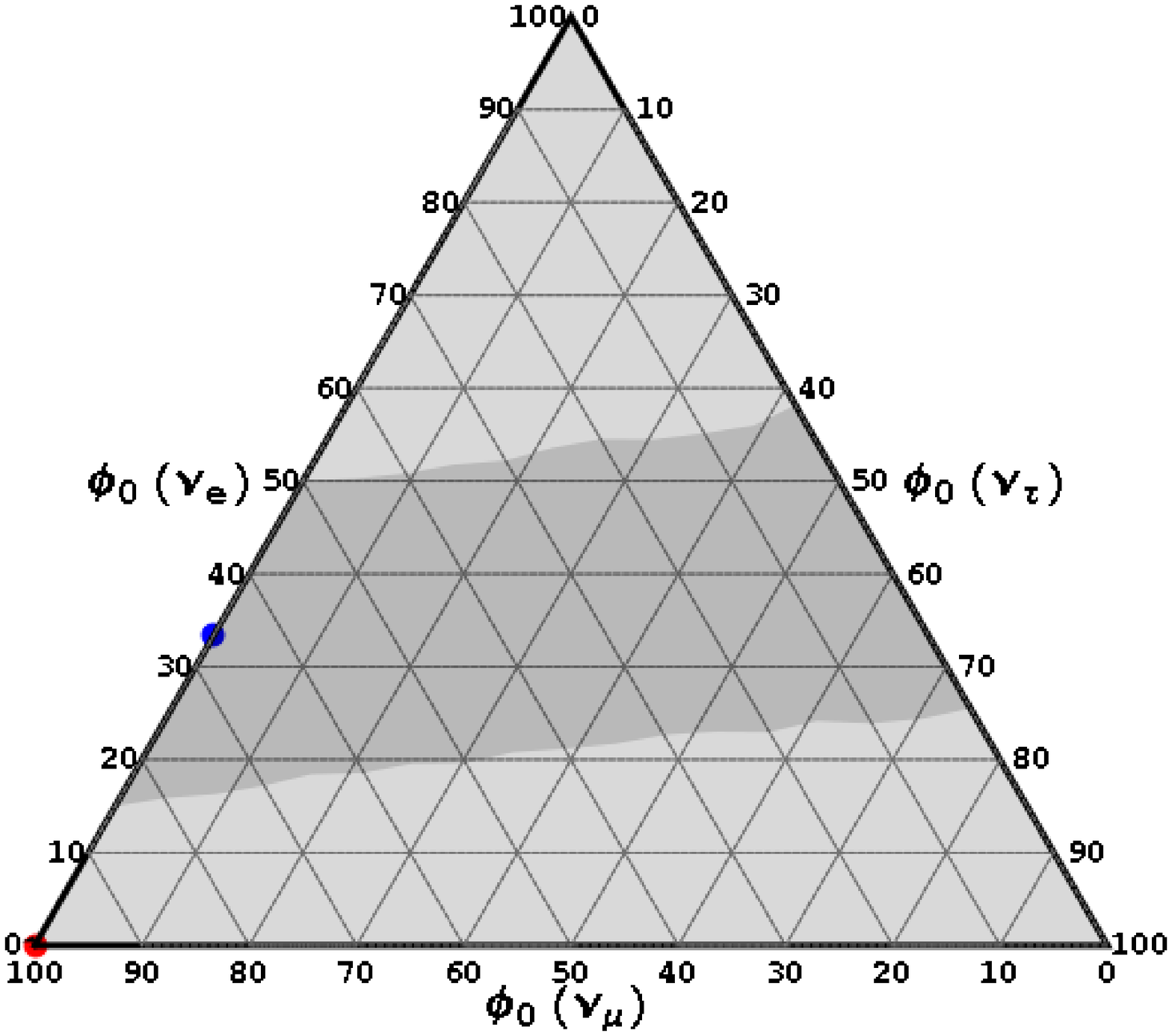}
\caption{The reconstructed ranges for the neutrino flavor ratios at
the source with $\Delta R_i/R_i=10\%$. The left and right panels are
results with the muon-damped source and the pion source as the input
true source respectively. The numbers on each side of the triangle
denote the flux percentage of a specific flavor of neutrino. The red
point marks the muon-damped source $\Phi_{0,\mu}=(0,1,0)$ and the
blue point marks the pion source $\Phi_{0,\pi}=(1/3,2/3,0)$. Gray
and light gray areas respectively denote the $1\sigma$ and $3\sigma$
ranges for the reconstructed neutrino flavor ratios at the source.
We choose parameter set 1 in Table~\ref{scen} for this analysis.}
\label{onlyR}
\end{figure}

It is instructive to see how well one can determine the initial
neutrino flavor ratio by measuring $R$ alone. We perform such an
analysis by neglecting the second term on the RHS of
Eq.~(\ref{chi}). The $1\sigma$ and $3\sigma$ ranges for the
reconstructed flavor ratios at the source are shown in
Fig.~\ref{onlyR}. For an input muon-damped source, it is seen that,
with $\Delta R_{\mu}/R_{\mu}=10\%$, the reconstructed $3\sigma$
range of the neutrino flavor ratio almost covers the entire physical
region. For an input pion source with $\Delta R_{\pi}/R_{\pi}=10\%$,
all possible initial neutrino flavor ratios are allowed at the
$3\sigma$ level. Clearly it is desirable to measure both $R$ and
$S$.

\subsection{The flavor reconstruction with measurements on both $R$ and $S$ }\label{un}
In this subsection, we perform a statistical analysis with respect
to simultaneous measurements of $R$ and $S$. The accuracy for the
measurement on $R$ is $\Delta R_i/R_i=10\%$ with $i=\pi, \, \mu$.
Here we adopt the first approach for estimating $\Delta S_i/S_i$
while present the second approach in the next subsection. With the
first approach, we have $\Delta S_{\pi}/S_{\pi}=(11-12)\%$ and
$\Delta S_{\mu}/S_{\mu}=(11-14)\%$ depending on the parameter set
chosen for calculations.

\subsubsection{$(\sin^2\theta_{13})_{\rm best \, fit}=0$}
\begin{figure}[htbp]
\includegraphics[scale=0.35]{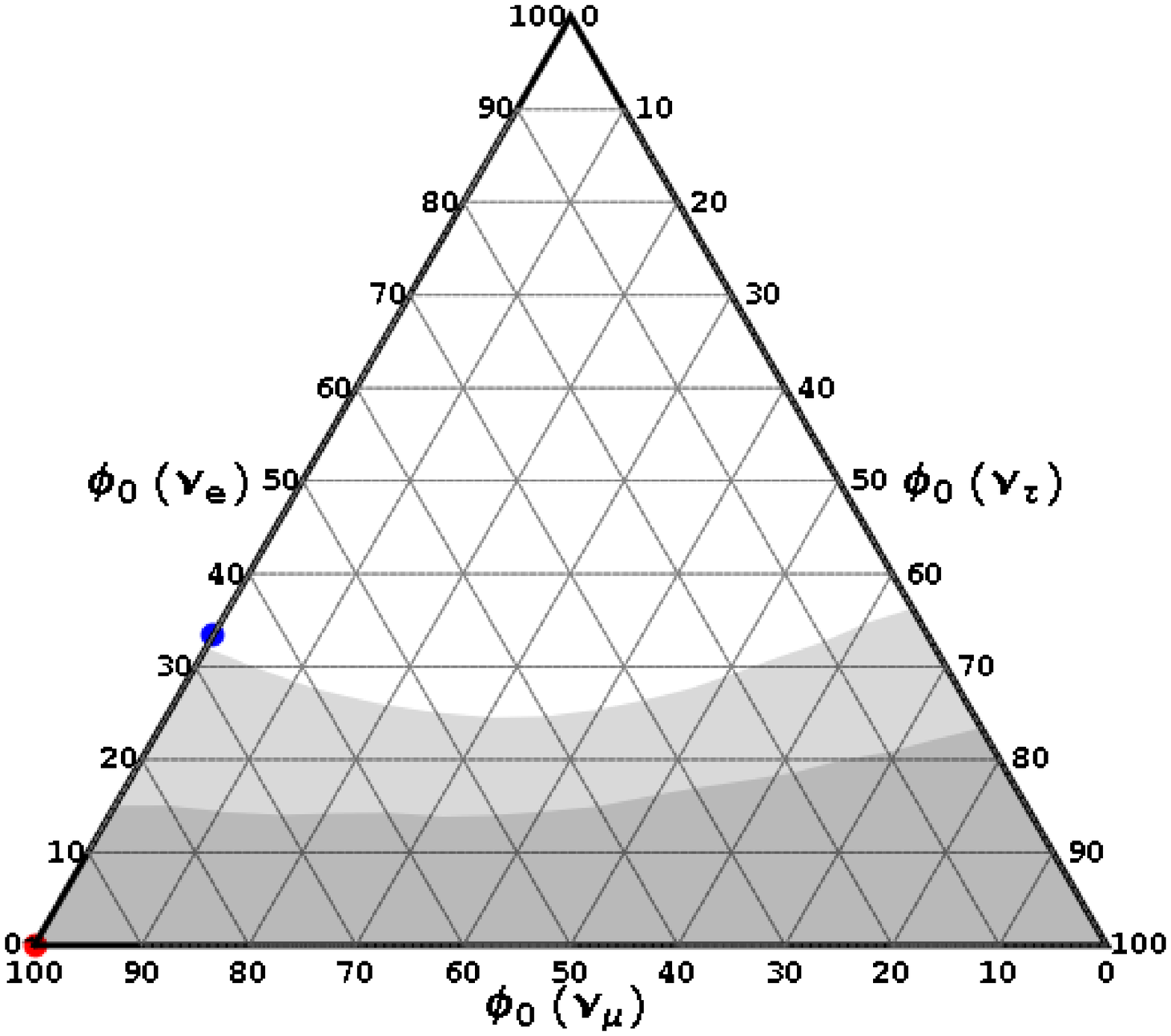}
\includegraphics[scale=0.35]{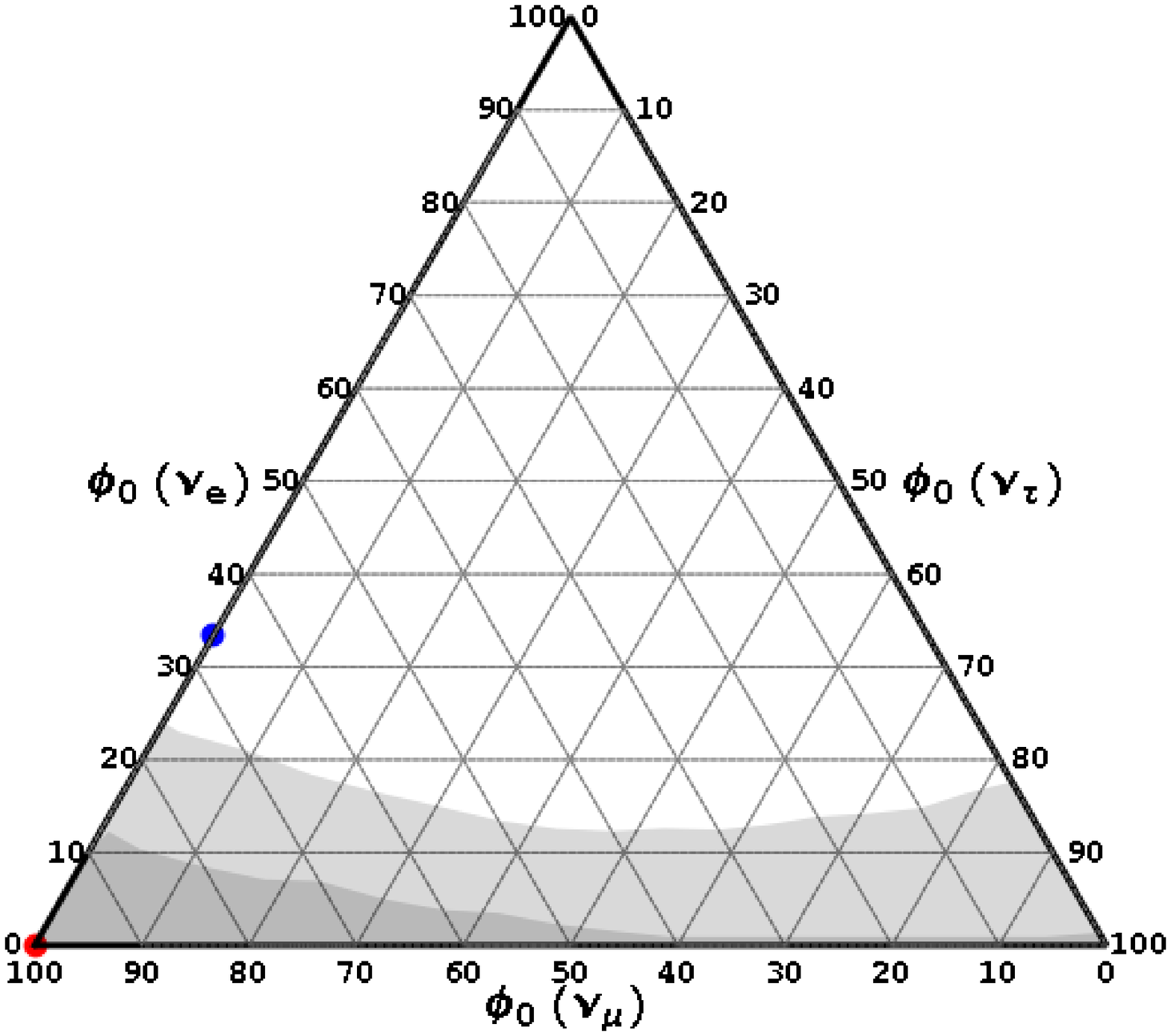}
\caption{The reconstructed ranges for the neutrino flavor ratios for an
input muon-damped source with $\Delta R_{\mu}/R_{\mu}=10\%$ and $\Delta S_{\mu}/S_{\mu}$
related to the former by the Poisson statistics.
Gray and light gray areas in the left (right) panel denote the reconstructed 1$\sigma$
and 3$\sigma$ ranges with the parameter set 1 (2).}
\label{mu1}
\end{figure}

\begin{figure}[htbp]
\includegraphics[scale=0.35]{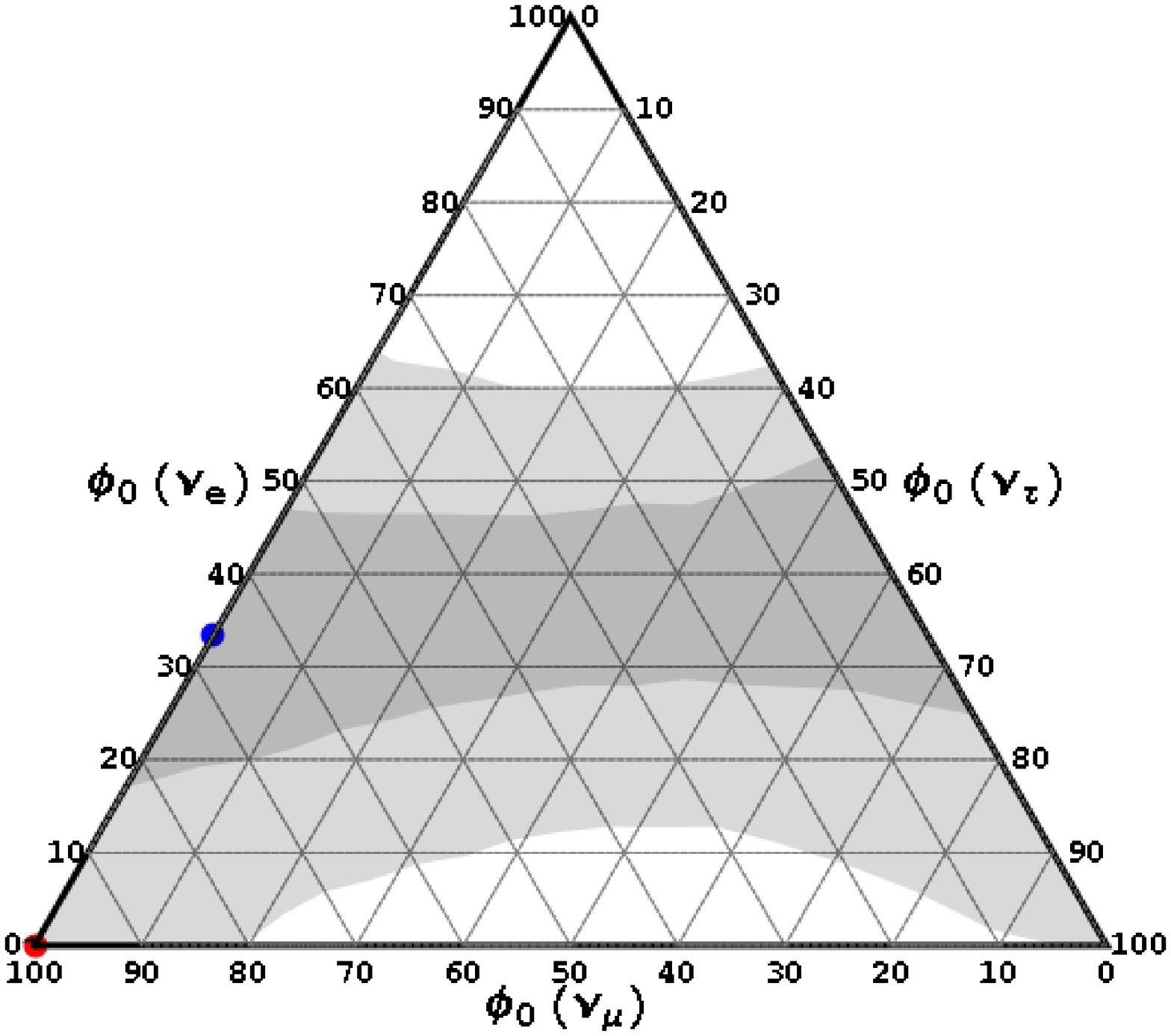}
\includegraphics[scale=0.35]{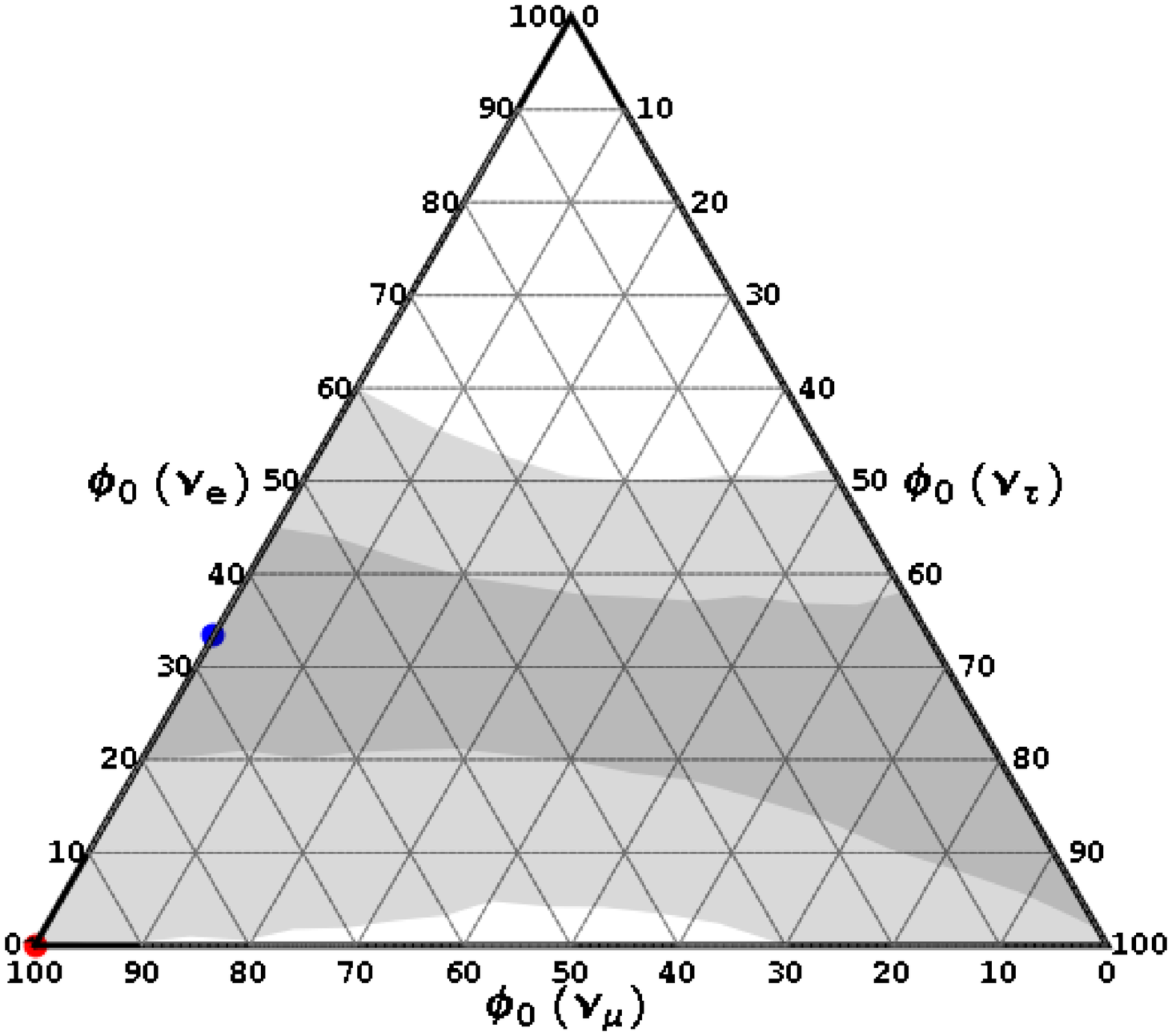}
\caption{ The reconstructed ranges for the neutrino flavor ratios for an input pion
source with $\Delta R_{\pi}/R_{\pi}=10\%$ and $\Delta S_{\pi}/S_{\pi}$ related to the former
by the Poisson statistics.
Gray and light gray areas in the left (right) panel denote the reconstructed
1$\sigma$ and 3$\sigma$ ranges with the parameter set 1 (2).}
\label{pi1}
\end{figure}

We begin our analysis with the parameter set 1 and 2 where
$(\sin^2\theta_{13})_{\rm best \, fit}=0$ and
$(\sin^2\theta_{23})_{\rm best \, fit}=0.45$ and $0.55$
respectively. Figs. \ref{mu1} and \ref{pi1} show the reconstructed
neutrino flavor ratios for an input muon-damped source and an input
pion source respectively. The reconstructed initial flavor ratios
are seen to include the region with significant $\nu_{\tau}$
fractions. It has been shown in Sec.~\ref{matrix} that the flux
combination $\left(1+4r\Delta\right)\phi_0 (\nu_{\mu})-
 \left(1-2r\Delta\right)\phi_0 (\nu_{\tau})$ is poorly
constrained due to the smallness of eigenvalue $\lambda^{\prime}_b$
associated with $V^{\prime b}$ (see Eq.~(\ref{lapabc}) and
(\ref{vpabc})). This then leads to an extension in the reconstructed
range of the initial neutrino flavor ratio along the $V^{\prime b}$
direction. In the limit $\Delta \equiv \cos 2\theta_{23}=0$, i.e.,
$\sin^2\theta_{23}=0.5$, $V^{\prime b}$ reduces to $V^b$ (see
Eq.~(\ref{vabc})) which is exactly parallel to the $\nu_e$-less side
of the flavor-ratio triangle. The direction of $V^{\prime b}$
deviates slightly from that of $V^b$ in opposite ways depending on
the sign of $\Delta$. This is seen by comparing the left and right
panels of both Fig. \ref{mu1} and Fig. \ref{pi1}. Due to
uncertainties of neutrino mixing parameters, we note that the
boundaries for $1\sigma$ and $3\sigma$ regions are not straight
lines. For an input muon-damped source, the pion source can be ruled
out at the $3\sigma$ level as shown in Fig.~\ref{mu1}. However, the
converse is not true as seen from Fig.~\ref{pi1}. Finally, as shown
in the right panel of Fig.~\ref{mu1}, an astrophysical hidden source
with $\Phi_{0, {\rm ah}}=(1/3, a, 2/3-a)$ \cite{Mena:2006eq,comment}
can be ruled out at the $3\sigma$ level for an input muon-damped
source with $(\sin^2\theta_{23})_{\rm best \, fit}=0.55$.

\subsubsection{$(\sin^2\theta_{13})_{\rm best \, fit}>0$}

\begin{figure}[htbp]
\includegraphics[scale=0.35]{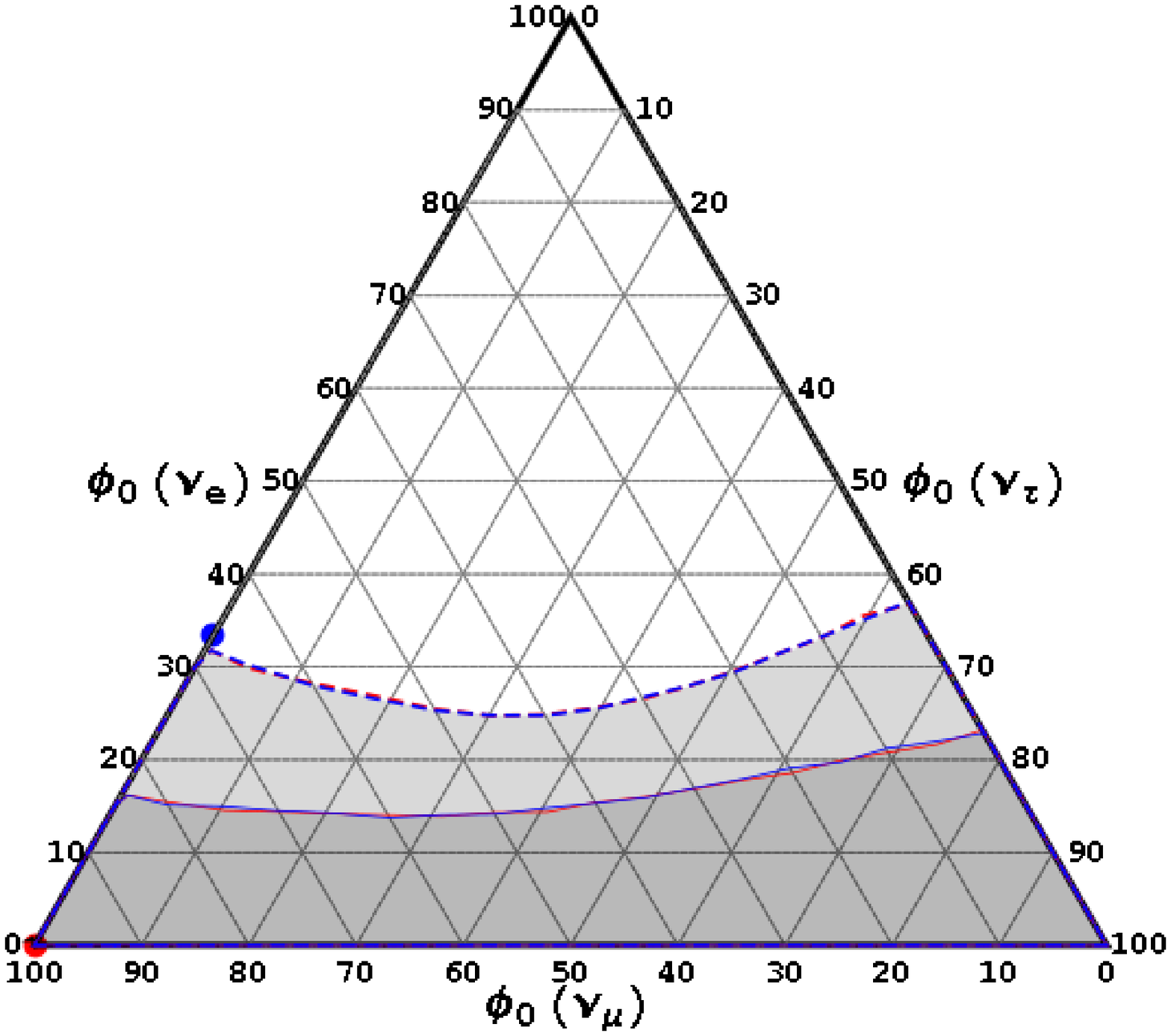}
\includegraphics[scale=0.35]{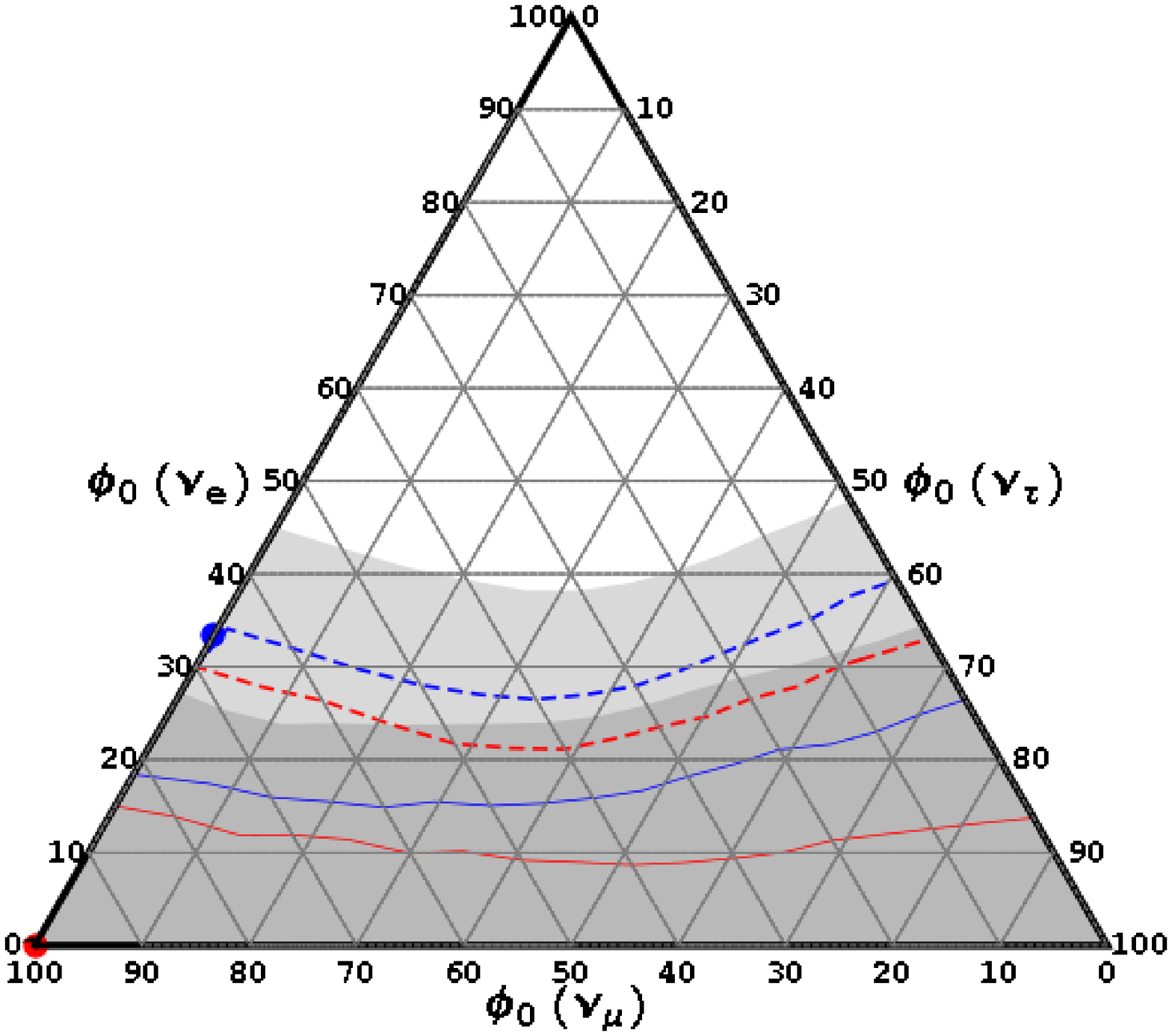}
\caption{ The reconstructed ranges for the neutrino flavor ratio at
the source for an input muon-damped source with $\Delta
R_{\mu}/R_{\mu}=10\%$ and $\Delta S_{\mu}/S_{\mu}$ related to the
former by the Poisson statistics. The left panel is obtained with
$\theta_{13}$ and $\theta_{23}$ taken from the parameter set 1 and
the input CP phase taken to be $0$, $\pi/2$ and $\pi$ respectively.
The right panel is obtained with the parameter sets 3a, 3b and 3c.
Light gray area, dashed blue and dashed red lines correspond to the
$3\sigma$ ranges for the reconstructed neutrino flavor ratio at the
source for $\cos\delta=1$, $\cos\delta=0$ and $\cos\delta=-1$
respectively. Gray area, blue and red lines correspond to the
$1\sigma$ ranges for the reconstructed neutrino flavor ratio at the
source for $\cos\delta=1$, $\cos\delta=0$ and $\cos\delta=-1$
respectively. The effect from the CP phase $\delta$ only appears in
the right panel.} \label{muCP}
\end{figure}

\begin{figure}[htbp]
\includegraphics[scale=0.36]{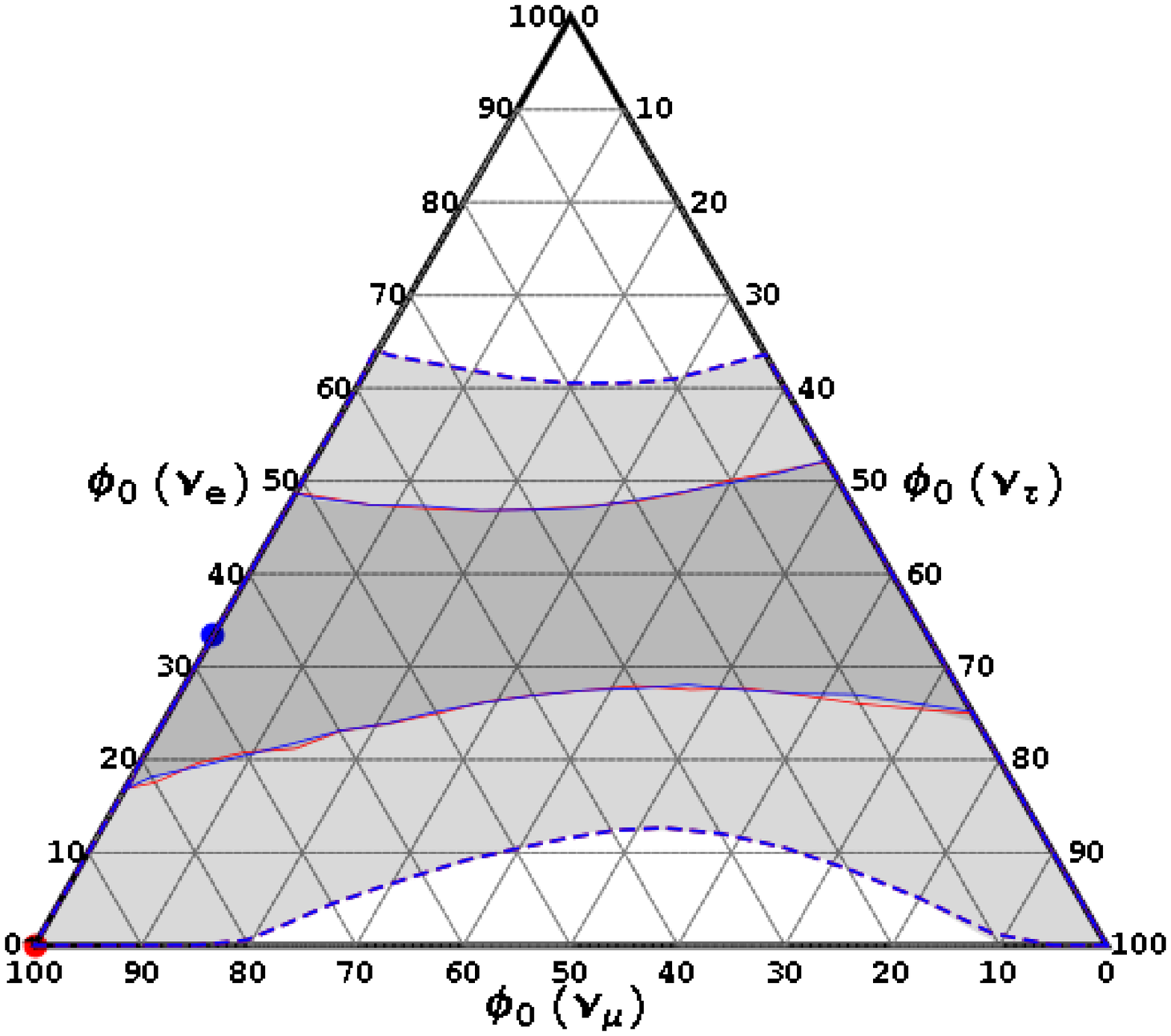}
\includegraphics[scale=0.36]{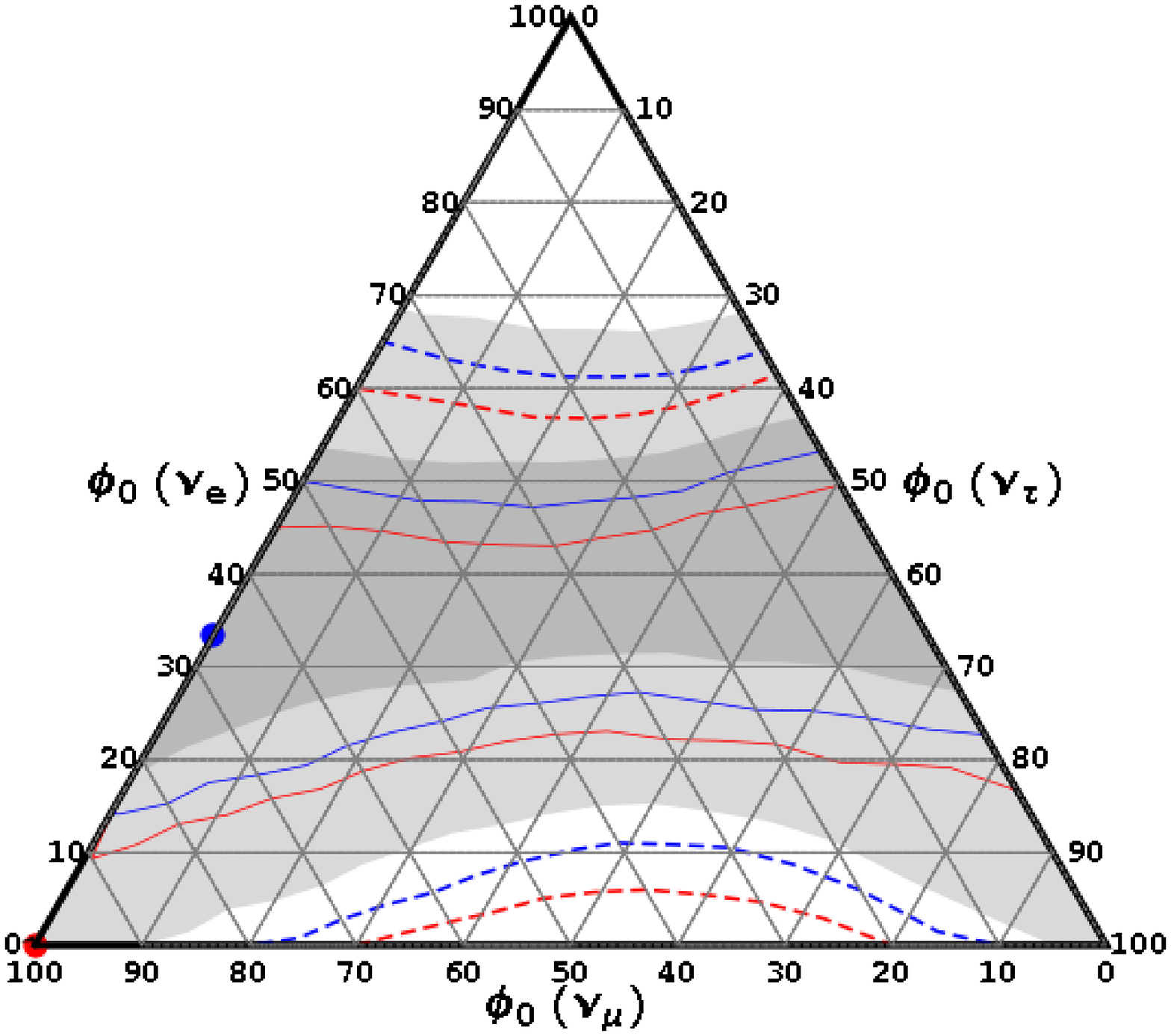}
\caption{The reconstructed $1\sigma$ and $3\sigma$ ranges for the
neutrino flavor ratio at the source for an input pion source with
$\Delta R_{\pi}/R_{\pi}=10\%$ and $\Delta S_{\pi}/S_{\pi}$ related
to the former by the Poisson statistics. The choices of parameter
sets are identical to those of Fig.~\ref{muCP}. Once more, the
effect from the CP phase $\delta$ only appears in the right panel.
}\label{piCP}
\end{figure}

A non-zero $\theta_{13}$ introduces the CP phase contribution to
every element of matrix $P$, except $P_{ee}$. We study the effect of
CP phase $\delta$ on the reconstruction of neutrino flavor ratio at
the source. We choose parameter sets 3a, 3b and 3c for performing
the statistical analysis. The results are shown in the right panels
of Figs. \ref{muCP} and \ref{piCP}. For comparisons, we also perform
the analysis with $\theta_{13}$ and $\theta_{23}$ taken from the
parameter set 1 and the input CP phase taken to be $0$, $\pi/2$ and
$\pi$ respectively. The results are shown in the left panels of
Figs. \ref{muCP} and \ref{piCP}.

Left panels of Figs. \ref{muCP} and \ref{piCP} indicate that the
reconstructed ranges for initial neutrino flavor ratios are
independent of the input CP phase for $(\sin^2\theta_{13})_{\rm best
\, fit}=0$. The dependencies on the CP phase only appear in the
right panels. For an input muon-damped source (see Fig.~\ref{muCP}),
the allowed $1\sigma$ and $3\sigma$ ranges for initial neutrino
flavor ratios are the smallest (denoted by red curves) for $\cos
\delta=-1$, i.e., $\delta=\pi$. In this case, the pion source and
the astrophysical hidden source mentioned earlier can both be ruled
out at the $3\sigma$ level \cite{comment}. The allowed ranges become
the largest (denoted by gray areas) for $\cos \delta=1$, i.e.,
$\delta=0$.  For an input pion source with different CP phases, the
allowed $3\sigma$ ranges for the initial neutrino flavor ratio
always cover the muon-damped source.

\subsection{Critical accuracies needed for distinguishing astrophysical
sources.}\label{crit} 
\begin{figure}[htbp]
\includegraphics[scale=0.35]{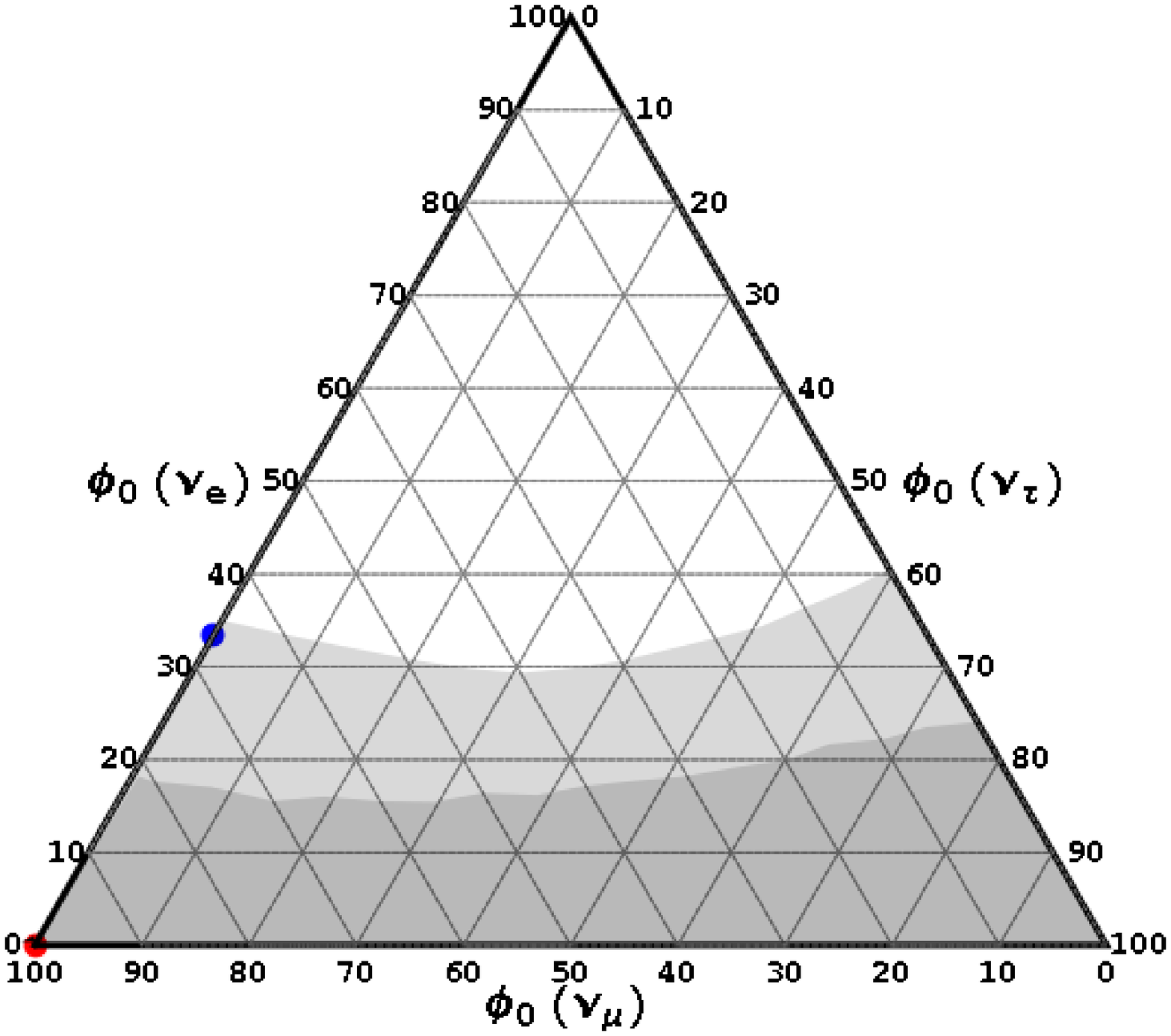}
\includegraphics[scale=0.35]{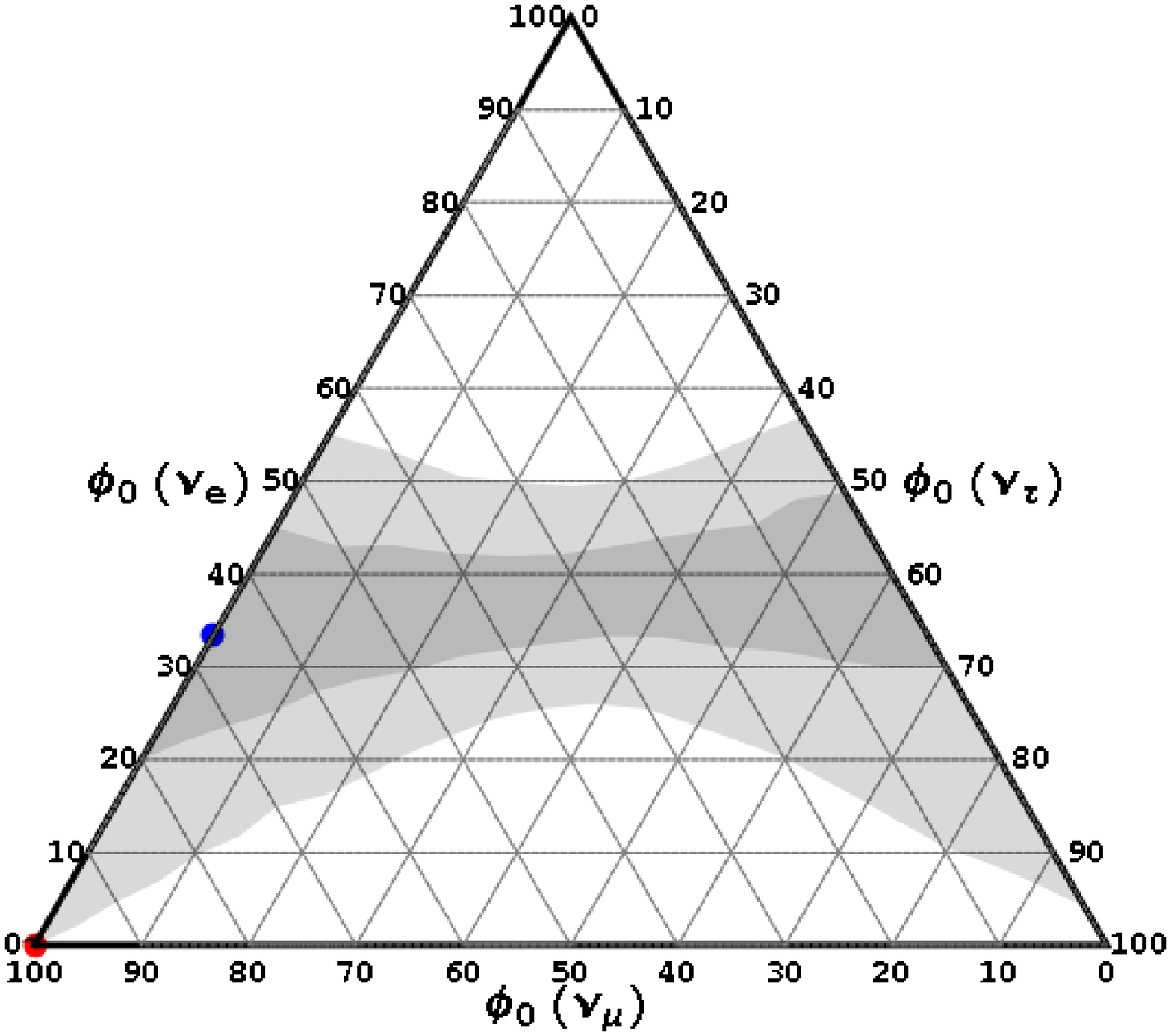}
\caption{Critical accuracies needed to distinguish between the pion
source and the muon-damped source. In the left panel where the
muon-damped source is the true source, the reconstructed $3\sigma$
range for the neutrino flavor ratio just touches the pion source at
$\Delta R_{\mu}/R_{\mu}=13\%$. In the right panel where the pion
source is the true source, the reconstructed $3\sigma$ range for the
neutrino flavor ratio just touches the muon-damped source at $\Delta
R_{\pi}/R_{\pi}=6\%$.  We choose parameter set 1 for this analysis.}
\label{distinguish_1}
\end{figure}

\begin{figure}[htbp]
\includegraphics[scale=0.35]{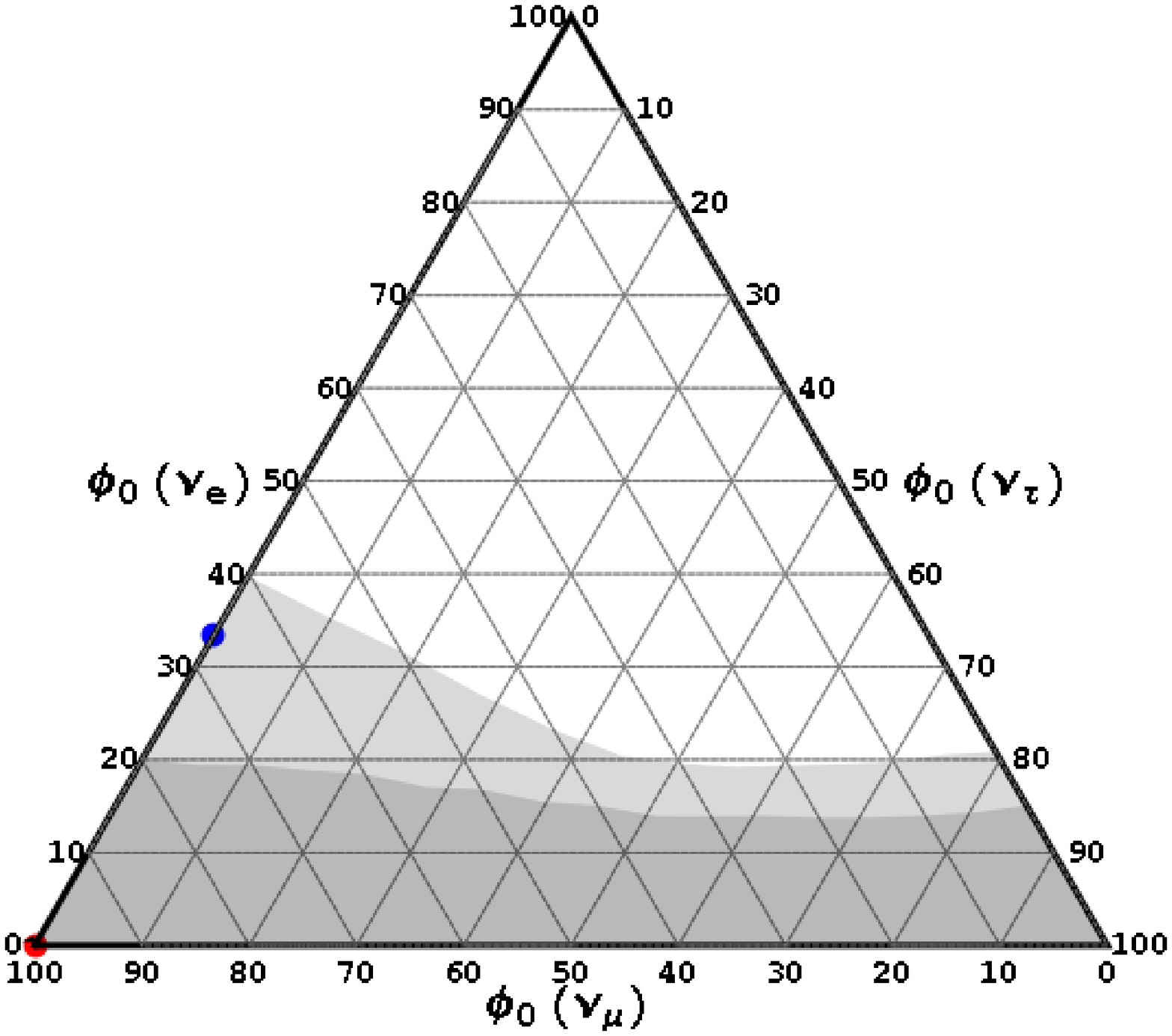}
\includegraphics[scale=0.35]{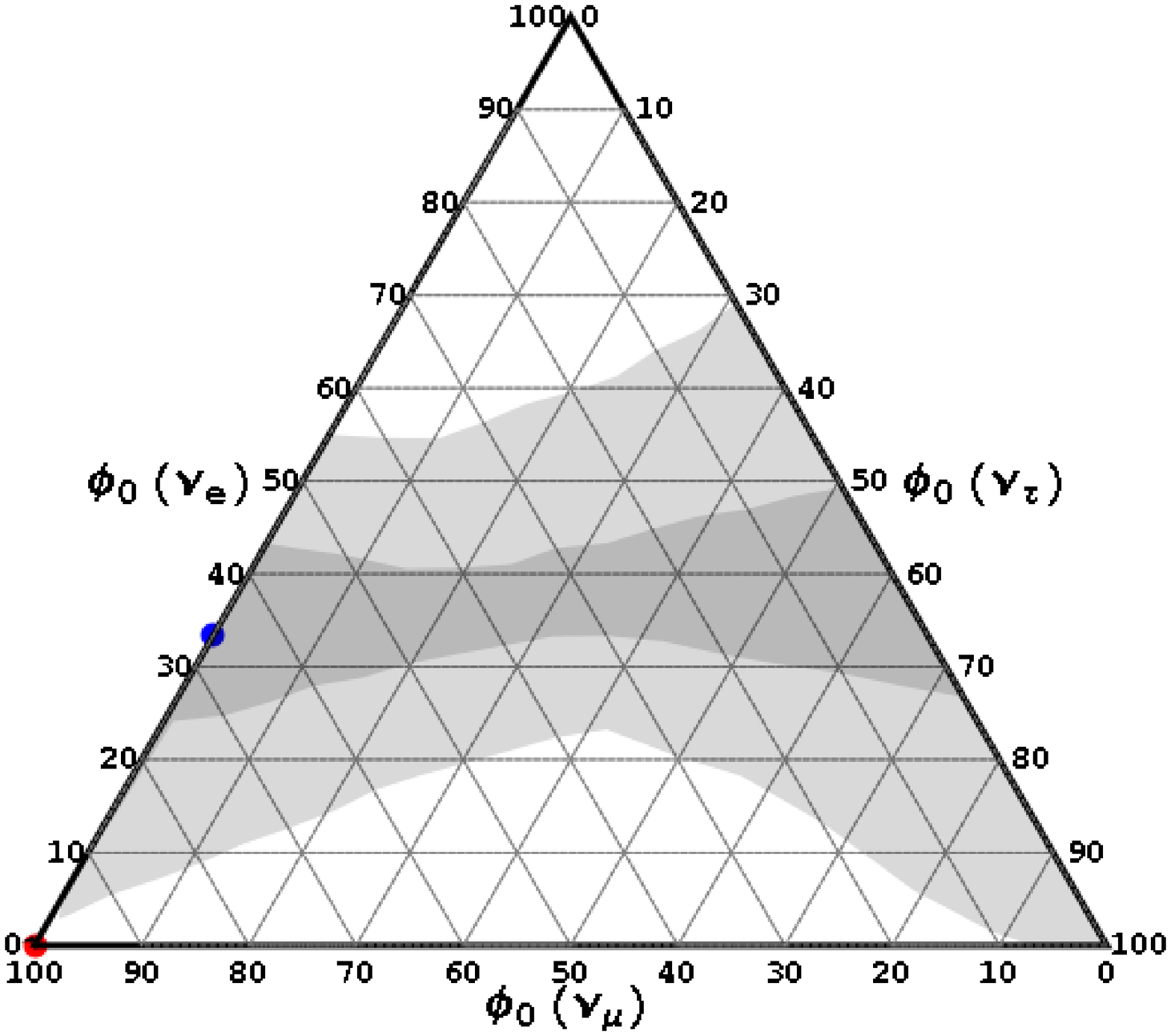}
\caption{Left panel: the reconstructed $1\sigma$ and $3\sigma$
ranges for the neutrino flavor ratio at the source for an input
muon-damped source with $\Delta S_{\mu}/S_{\mu}=25\%$ and $\Delta
R_{\mu}/R_{\mu}=2\%$. Right panel: the reconstructed $1\sigma$ and
$3\sigma$ ranges for the neutrino flavor ratio at the source for an
input pion source with $\Delta S_{\pi}/S_{\pi}=15\%$ and $\Delta
R_{\pi}/R_{\pi}=1.5\%$. We choose parameter set 1 for this
analysis.} \label{distinguish_2}
\end{figure}

It is important to identify critical
accuracies in measurement needed to distinguish between the pion
source and the muon-damped source. Choosing the parameter set 1 for
the analysis, we present the results in Figs. \ref{distinguish_1}
and \ref{distinguish_2} where two different approaches for
determining $\Delta S_i/S_i$ are used.
In Fig.~\ref{distinguish_1}, we determine $\Delta S_i/S_i$ by
applying Poisson statistics. In the left panel of
Fig.~\ref{distinguish_1}, which has the muon-damped source as the
true source, the reconstructed $3\sigma$ range for the neutrino
flavor ratio just touches the pion source at $\Delta
R_{\mu}/R_{\mu}=13\%$ and $\Delta S_{\mu}/S_{\mu}=16\%$. In the
right panel of this figure, which has the pion source as the true
source, the reconstructed $3\sigma$ range for the neutrino flavor
ratio just touches the muon-damped source at $\Delta
R_{\pi}/R_{\pi}=6\%$ and $\Delta S_{\pi}/S_{\pi}=7\%$. In
Fig.~\ref{distinguish_2}, we fix $\Delta S_{\mu}/S_{\mu}=25\%$ for
the left panel and fix $\Delta S_{\pi}/S_{\pi}=15\%$ for the right
panel. The result in the left panel is for $\Delta
R_{\mu}/R_{\mu}=2\%$. We find that the pion source can be ruled out
at the $3\sigma$ level if $\Delta R_{\mu}/R_{\mu}$ is lowered to
$1\%$. The result in the right panel is for $\Delta
R_{\pi}/R_{\pi}=1.5\%$. If $\Delta R_{\pi}/R_{\pi}$ is raised to
$2\%$, we find that the muon-damped source can not be ruled out at
the $3\sigma$ level. We have also investigated the case $\Delta
S_{\pi}/S_{\pi}=25\%$. In this case the reconstructed $3\sigma$
range of the neutrino flavor ratio covers the entire physical region
unless $\Delta R_{\pi}/R_{\pi}$ is smaller than $1\%$.

\section{Discussion and Conclusion}
The structure of the oscillation probability matrix $P$ (singular in
the limit $\theta_{23}=\pi/4$ and $\theta_{13}=0$) makes it
difficult to constrain a flux combination approximately like the
difference between $\phi_0(\nu_{\mu})$ and $\phi_0(\nu_{\tau})$.
This then leads to an extension in the reconstructed range for the
initial neutrino flavor ratio along the direction of $V^{\prime b}$.

We have illustrated the reconstruction of the neutrino flavor ratio
at the source from the measurements of energy-independent ratios
$R\equiv\phi (\nu_{\mu})/\left(\phi (\nu_{e})+\phi
(\nu_{\tau})\right)$ and $S\equiv\phi (\nu_e)/\phi (\nu_{\tau})$
among integrated neutrino flux. The ranges of neutrino mixing
parameters used in this analysis are summarized in
Eq.~(\ref{bestfit}). By just measuring $R$ alone from either an
input pion source or an input muon-damped source with a precision
$\Delta R/R=10\%$, the reconstructed $3\sigma$ range for the initial
neutrino flavor ratio is almost as large as the entire physical
range for the above ratio. By measuring both $R$ and $S$ from an
input muon-damped source, the pion source can be ruled out at the
$3\sigma$ level for the parameter sets 1 and 2 with $\Delta
R_{\mu}/R_{\mu}=10\%$ and $\Delta S_{\mu}/S_{\mu}$ related to the
former by the Poisson statistics. With a pion source as the input
true source and the choice of parameter set 1 for our analysis, the
muon-damped source can not be ruled out at the $3\sigma$ level until
$\Delta R_{\pi}/R_{\pi}$ and $\Delta S_{\pi}/S_{\pi}$ reach to $6\%$
and $7\%$ respectively. In the case $(\sin^2\theta_{13})_{\rm best
\, fit}>0$ as suggested by Ref.~\cite{Fogli:2008jx}, the CP phase
$\delta$ is seen to affect the reconstructed range for the neutrino
flavor ratio at the source. We have also presented results for
$\Delta S_{\mu}/S_{\mu}$ and $\Delta S_{\pi}/S_{\pi}$ fixed at
$25\%$ and $15\%$ respectively. To distinguish the pion source and
the muon-damped source in this case, both $\Delta R_{\mu}/R_{\mu}$
and $\Delta R_{\pi}/R_{\pi}$ should be of the order $1\%$ or
smaller.

We have also performed a statistical analysis with the errors of
$\theta_{23}$ and $\theta_{12}$ both reduced and the limit of
$\theta_{13}$ improved to $\sin^2\theta_{13}< 0.0025$ (i.e., $\sin^2
2\theta_{13}< 0.01$). The result of this analysis can be best
described by the modification to the left panel of Fig.~\ref{pi1}.
With $\sin^2\theta_{13}< 0.0025$, it is possible to rule out the
muon-damped source at the $3\sigma$ level for an input pion source
by reducing the errors of both $\theta_{23}$ and $\theta_{12}$ to
$70\%$ of their original values.

We like to point out that our analysis has been based upon the ideal
scenario that the true astrophysical neutrino source is either a
pure pion source or a pure muon-damped source. In practice, the
neutrino flavor ratio in a single astrophysical source can depend on
neutrino energies such that it behaves like the one from a pion
source at the low energy and gradually makes a transition to the one
from a muon-damped source as the neutrino energy increases
\cite{Kashti:2005qa}. Hence the reconstruction of the source flavor
ratio ought to be carried out separately for low and high energy
portions of the data. Furthermore an analysis on the diffuse
neutrino flux is challenging since such a flux arises from
astrophysical sources with different neutrino flavor ratios.
Nevertheless, the very high energy part of the flux spectrum is
possibly dominated by the cosmogenic neutrino flux \cite{cosmogenic}
arising from GZK \cite{Greisen,ZK} interactions. The cosmogenic
neutrino flux is a typical example of neutrino fluxes due to the
pion source. Therefore it is sensible to reconstruct the source
flavor ratio with respect to the highest energy part of the diffuse
neutrino spectrum, provided there are sufficient number of events.

The accuracy $\Delta R_i/R_i=10\%$ ($i=\mu, \, \pi$) frequently used
in our discussions requires $\mathcal{O}(100)$ neutrino events for
each flavor. Furthermore, the above accuracy requires an improved
understanding on the background atmospheric neutrino flux to a level
better than $10\%$ in the future. Taking a neutrino flux upper bound
$E^2\phi (\nu_{\alpha})= 10^{-7}$ GeV cm$^{-2}$s$^{-1}$ ($\alpha=e,
\, \mu, \tau$) derived by Waxman and Bahcall \cite{Waxman:1998yy},
we estimate by a simple re-scaling of the result in
Ref.~\cite{Ahrens:2003ix} that it takes the IceCube detector about a
decade to accumulate $\mathcal{O}(100)$ $\nu_{\mu}$ events. We
stress that the above bound is for diffuse neutrino flux. The flux
from individual point source is smaller. Hence it takes even a
longer period to accumulate the same number of events.
 The IceRay \cite{Allison:2009rz} detector is expected to
accumulate neutrino events in a much faster pace. However the
efficiency of flavor identification in this detector still requires
further studies.

In summary, we have demonstrated that it is challenging to
reconstruct the neutrino flavor ratio at the astrophysical source,
requiring a lot more than a decade of data taking in a neutrino
telescope such as IceCube for distinguishing between the pion source
and the muon-damped source. We stress that the large uncertainty in
the flavor ratios of astrophysical neutrinos should be taken into
account as one uses these neutrinos as a beam source to extract the
neutrino mixing parameters.

{\it Note added.} As we were writing up this paper, we became aware of a paper by A. Esmaili and Y. Farzan, arXiv:0905.0259 [hep-ph], which also discusses the initial flavor composition of cosmic neutrinos with an approach different from ours.

\noindent{\bf Acknowledgements}
This work is supported by National Science Council of
Taiwan under the grant numbers 97-2811-M-009-029 and 96-2112-M-009-023-MY3.

\appendix

\section{The exact oscillation probabilities of astrophysical neutrinos}
The exact analytic expressions for the components of $P$ are given by
\begin{eqnarray}
P_{ee}&=&\left(1-\frac{1}{2}\omega\right)(1-D^2)^2+D^4,\nonumber \\
P_{e\mu}&=&\frac{1}{4}(1-D^2)\left[\omega(1+\Delta)+(4-\omega)
(1-\Delta)D^2+2\sqrt{\omega(1-\omega)(1-\Delta^2)}D\cos\delta\right],\nonumber
\\
P_{e\tau}&=&\frac{1}{4}(1-D^2)\left[\omega(1-\Delta)+(4-\omega)
(1+\Delta)D^2-2\sqrt{\omega(1-\omega)(1-\Delta^2)}D\cos\delta\right],\nonumber
\\
P_{\mu\mu}&=&\frac{1}{2}\left[(1+\Delta^2)-(1-\Delta)^2D^2(1-D^2)\right]\nonumber
\\
&-&\frac{1}{8}\omega\left[(1+\Delta)^2+(1-\Delta)^2D^4-(1-\Delta^2)
D^2(2+4\cos^2\delta)\right]\nonumber
\\
&-&\frac{1}{2}\sqrt{\omega(1-\omega)(1-\Delta^2)}
\left[(1+\Delta)-(1-\Delta)D^2\right]D\cos\delta,\nonumber \\
P_{\mu\tau}&=&\frac{1}{2}(1-\Delta^2)(1-D^2+D^4)\nonumber \\
&-&\frac{1}{8}\omega\left[(1-\Delta^2)(1+4D^2\cos^2\delta+D^4)-2(1+\Delta^2)D^2\right]\nonumber \\
&+&\frac{1}{2}\sqrt{\omega(1-\omega)(1-\Delta^2)}\Delta (1+D^2)D\cos\delta,\nonumber \\
P_{\tau\tau}&=&\frac{1}{2}\left[(1+\Delta^2)-(1+\Delta)^2D^2(1-D^2)\right]\nonumber \\
&-&\frac{1}{8}\omega\left[(1-\Delta)^2+(1+\Delta)^2D^4-(1-\Delta^2)
D^2(2+4\cos^2\delta)\right]\nonumber \\
&+&\frac{1}{2}\sqrt{\omega(1-\omega)(1-\Delta^2)}
\left[(1-\Delta)-(1+\Delta)D^2\right]D\cos\delta,
\label{exact_prob}
\end{eqnarray}
where $\omega\equiv \sin^2 2\theta_{12}$, $\Delta\equiv \cos 2\theta_{23}$, $D\equiv \sin\theta_{13}$, and $\delta$ the CP phase.

\end{document}